\documentstyle[12pt]{article}
\newcommand{\beq}{\begin{equation}}
\newcommand{\eeq}{\end{equation}}

\newcommand{\ba}{\begin{array}}
\newcommand{\ea}{\end{array}}

\newcommand{\dsp}{\displaystyle}
\newcommand{\EQN}{\label}

\begin{document}

\begin{titlepage}
\noindent
%
%
\hfill TTP94--7\\
\mbox{}
\hfill May 1994   \\   
%
%
\vspace{0.5cm}
\begin{center}
  \begin{Large}
  \begin{bf}
Corrections of Order
 ${\cal O}(G_F\alpha_s m_t^2)$
 to the Higgs Decay Rate
 $\Gamma(H\to b\bar{b})$
  \end{bf}
  \end{Large}\footnote[1]{A complete postscript file,
including figures, is available via anonymous ftp
at ttpux2.physik.uni-karlsruhe.de (129.13.102.139)
as /ttp94-07/ttp94-07.ps}
%
%

  \vspace{0.8cm}
  \begin{large}
 A.Kwiatkowski\footnote[2]{Present address:
   \begin{minipage}[t]{8cm}
              Lawrence Berkeley Laboratory \\
              Bldg. 50 A - 3115, \\
              1 Cyclotron Road \\
              Berkeley, CA. 94720, USA
   \end{minipage}     }
 M.Steinhauser    \\[3mm]
    Institut f\"ur Theoretische
Teilchenphysik\\
    Universit\"at Karlsruhe\\
    Kaiserstr. 12,
  Postfach 6980\\[2mm]
   D-76128 Karlsruhe, Germany\\
  \end{large}

%
%
  \vspace{3cm}
  {\bf Abstract}
\end{center}
\begin{quotation}
\noindent
QCD corrections to the electroweak
 one-loop result for the partial width
$\Gamma(H\rightarrow b\bar{b})$ are
studied. For the decay channel into
 bottom quarks the rate is affected by
a virtual top quark through electroweak
interactions. The calculation of QCD
corrections to this quantity is performed
for an intermediate range Higgs mass in
the heavy top mass limit. The leading
correction
of order ${\cal O}(G_F\alpha_s m_t^2)$
is estimated. Numerically the contribution
is of comparable
 size as the electroweak
correction, but of opposite sign.
\end{quotation}

\end{titlepage}

\renewcommand{\arraystretch}{2}

\section{Introduction}
At present high energy colliders experimental
data have turned out to be in remarkable
agreement with theoretical predictions
of the Standard Model. Precision experiments
have covered many aspects of both its
electroweak and  QCD sector.
Although the formulation of the Standard
Model as a $SU(3)_C\times
SU(2)_L\times U(1)_Y$ gauge theory is firmly
established,
experimental evidence
for the physical Higgs boson is still
missing. Since the Higgs mechanism of
spontaneous
symmetry breaking and mass generation
would reveal its structure through the
properties of the Higgs boson, the high mass $M_H$
of this scalar particle hides
the  Higgs sector
completely from observation at present
 energies.
Future accelerators like LHC and NLC,
which  are dedicated to the search
and the
study  of the Higgs particle, may
hopefully close the
energy gap to its production threshold.

For intermediate Higgs masses $M_H<2M_W$
the dominant decay channel is the decay
into bottom quarks.
A theoretical prediction for the partial
width $\Gamma(H\rightarrow b\bar{b})$
is therefore needed for the physics analysis
with high precision.
Consequently  much work has been spent
 in the past on the calculation of
radiative corrections to this
quantity. Excellent reviews on the  Higgs
phenomenology can be found for example
in \cite{GunHabKanDaw90,Kni93}.

Electroweak radiative corrections,
 which were studied at the one-loop level
 by several
 groups \cite{BarVilKhr91,Kni92,DabHol92},
are  of particular interest for
the process $H\rightarrow b\bar{b}$,
since they are affected by the mass of the
top quark.
Virtual top states are possible due to
bottom-top transitions mediated by
charged Higgs ghosts
$\Phi^{\pm}$ or $W^{\pm}$ bosons. The
 leading $m_t^2$ dependence originates from
diagrams with an exchange of $\Phi^{\pm}$
due to
the Yukawa coupling of the Higgs ghosts to
the fermion line.

In this work QCD corrections
to the electroweak one-loop result
 are calculated in the heavy top
 mass limit.
For this purpose the
  hard mass procedure
 \cite{PivTka84,GorLar87,CheSmi87,Smi91}
is employed. It results in a power series
in the inverse heavy mass $1/m_t$, of which
we compute the leading term of order
 ${\cal O}(G_F\alpha_s m_t^2)$.

It is convenient to consider the
corresponding scalar current correlator
$\Pi(q^2)$, since its
 absorptive part completely
determines the partial decay rate
\beq \EQN{i1}
\Gamma(H\to b\bar{b})=\frac{1}{M_H}\mbox{Im}\Pi(M_H^2)
.\eeq
The calculation of $\Pi(q^2)$ involves
multiloop propagator type integrals as well as
massive tadpole integrals.
Many of the
computational tools, mostly based on the
algebraic manipulation language FORM
\cite{Ver91}, have already been used
previously for the
calculation of
${\cal O}(G_F\alpha_sm_t^2)$ corrections
to the partial width $\Gamma(Z\rightarrow
b\bar{b})$ of the $Z$ boson
 \cite{CheKwiSte93}.
Nevertheless, the structures of the
calculations for the processes
$H\rightarrow b\bar{b}$ and
$Z\rightarrow b\bar{b}$ differ in
various respects.
First, the generic diagram with a primordial
decay of the Higgs into a pair of charged
Higgs ghosts and their
 subsequent transition
into b quarks does not contribute to
the considered order. This can be seen
 from the  $H\Phi^+\Phi^-$ coupling
 and is explained on dimensional grounds.
 Second, the Yukawa
couplings of the Higgs to the
scalar fermion currents are proportional
to the quark masses. Combined with
mass terms from the fermion traces
 the decay rate is proportional to the
square of the bottom mass
 and would vanish in the limit $m_b=0$.
Third, the quadratic dependence of the
correction term in the considered order
on the bottom as well as the top mass
affects the
renormalization of the scalar correlator.
Wheras for the $Z$ decay the sum of
the three loop diagrams was finite,
this is not the case for the Higgs rate.
The renormalization of the quark masses
induces contributions from lower order
diagrams which have to be included to
arrive at a finite result.

In Section 2.1 the calculation
of the three loop propagator diagrams is
described.
Subsequently the relation between
renormalized and bare quark masses
including  corrections up to the two loop
level is combined with the Born graph
as well as  pure electroweak and QCD
diagrams. The corresponding contributions
to the order
${\cal O}(G_F\alpha_sm_t^2)$ corrections
are discussed in Section 2.2.
In Section 3 universal corrections to
$\Gamma(H\rightarrow b\bar{b})$ are added
to the flavour specific corrections.
The numerical size of the effects is
evaluated.

\begin {figure}
\begin {center}
\begin {tabular}{cc}
\end {tabular}
\end {center}
\caption {The electroweak diagrams
 contributing to
the self energy of the $H$ boson.
Internal dashed line: Higgs ghost,
thin lines: bottom quark, thick lines:
top quark.}
\end{figure}

\section{Description of the Calculation}
\subsection{Three-Loop  Diagrams}
All diagrams of order
${\cal O}(G_F\alpha_sm_t^2)$ are obtained from
two different generic graphs (see Figure 1),
to which a gluon has to be attached in all
possible ways. The calculation is performed
with dimensional regularisation
in the on-shell scheme. In
$D=4-2\epsilon$ dimensions the definition
of the Hermitian, anticommuting
$\gamma_5$ matrix is used. After combining
pairs of $\gamma_5$ to unity, traces with
single $\gamma_5$ vanish identically.
 We work in the t'Hooft gauge
with an arbitary QCD gauge
 parameter $\xi_S$.
The verification of gauge
invariance serves as
an internal check of our result.

Being interested in the leading correction
of the power series with respect to the
inverse heavy mass $1/m_t$, we apply the
hard mass procedure as developed and
elaborated in
 \cite{PivTka84,GorLar87,CheSmi87,Smi91}.
Those subgraphs are selected, which
comprise all top propagators and become
one-\-particle-\-irreducible after
contracting the heavy lines to a point.
They are expanded with respect to
external  momenta and small masses.
This formal Taylor expansion is then
inserted as an effective vertex into
the remaining diagram which has to be
evaluated. In the heavy top limit the
 bottom mass as well as the
$W$ mass and the Higgs mass
 are considered small as compared
to the top mass.
Since the expansion is performed only to the
first nonvanishing order, the Higgs ghost
propagators are always contained in the
hard subgraphs. For the leading term
$M_W$ is therefore set to zero. As a
consequence the
electroweak gauge parameter drops out trivially.
Although the bottom mass
is also a small expansion parameter, it
 cannot be neglected throughout, since that
would result in an identically vanishing
decay rate. The bottom mass is not only
introduced as an overall factor through
the Yukawa coupling, but is also present
in  fermion traces with an odd number
of Dirac matrices for $m_b=0$.
Besides the expansion of the hard
 subgraphs as prescribed by the hard mass
procedure also the remaining diagram is
expanded with respect to $m_b^2$.
After all expansions are performed,
the
 leading piece can be isolated. Its
contribution $\Delta\Gamma_{\alpha\alpha_s}$
 to the partial Higgs decay rate reads
\beq\EQN{t1}
\ba{ll}\dsp
\Delta\Gamma_{\alpha\alpha_s}
& \dsp
=\Gamma_0m_b^2
 x_t\frac{\alpha_s}{\pi}
\left\{
9\frac{1}{\epsilon^2}
+\frac{1}{\epsilon}
\left(
19+12\ln\frac{\mu^2}{m_t^2}
-15\ln\frac{M_H^2}{\mu^2}
\right)
\right.
\\ & \dsp
+\frac{199}{3}-24\zeta(3)
-\frac{23}{6}\pi^2
+16\ln\frac{\mu^2}{m_t^2}
-41\ln\frac{M_H^2}{\mu^2}
\\ & \dsp
\left.
+9\ln^2\frac{\mu^2}{m_t^2}
+\frac{27}{2}\ln^2\frac{M_H^2}{\mu^2}
-18\ln\frac{\mu^2}{m_t^2}
\ln\frac{M_H^2}{\mu^2}
\right\}
\ea
\eeq
with
\beq\EQN{t2}
\ba{ll}\dsp
\Gamma_0
&\dsp
=  \frac{G_F}{4\sqrt{2}\pi}N_CM_H
\\ \dsp
 x_t
& \dsp
=\frac{G_Fm_t^2}{8\sqrt{2}\pi^2}
.\ea\eeq
According to the conventions of
the multiloop integration package
MINCER \cite{LarTkaVer91}, which we used for
the calculation of massless integrals,
terms with $\gamma_E$ and $\ln(4\pi)$
are suppressed.
In addition for each loop integration
a factor $e^{\zeta(2)\epsilon^2/2}$
is included for convenience.
It is clear, that physical results are not
influenced by this convention, since
only terms
of order $\epsilon^2$ are affected.
A consistent convention holds also
for the calculation of massive
tadpole integrals \cite{Che93}.
Our  result is obtained in the
limit of large top masses and is
applicable only in the range of intermediate
Higgs masses, in particular in the regime
$M_H<2m_t$ below the top threshold.

\subsection{Mass Renormalization and
Induced Contributions}
The result of eq.(\ref{t1})
is still divergent. Renormalization of the
quark masses induces additional
contributions which lead to a finite
expression for the Higgs decay rate.
The relation between the bare mass $m_b$
and the renormalized mass $m_b^{OS}$
of the bottom quark in the on-shell scheme
may be written as follows
\beq\EQN{m1}
m_b^{OS} = m_b\left\{
1+C_{\alpha}x_t+C_{\alpha_s}
\frac{\alpha_s}{\pi}
+C_{\alpha\alpha_s}
x_t\frac{\alpha_s}{\pi}
\right\}
\eeq
where only corrections of orders relevant
for our problem are taken into account.
The
corresponding coefficients
 $C_{\alpha},C_{\alpha_s}$ and
$C_{\alpha\alpha_s}$ can be derived
from the fact that
the renormalized (pole) mass
$m_b^{OS}$ is defined
through the location of the pole of the
quark
propagator
\beq\EQN{m2}
S_F=\frac{1}{i}\frac{1}{m_b-p\hspace{-.45em}/+\Sigma(p)}
.\eeq
 The selfenergy  $\Sigma(p)$
of the bottom quark is decomposed
in the form
\beq\EQN{m3}
\Sigma(p)=m_b\Sigma_2
 \left( \frac{m_b^2}{p^2} \right)
         +(m_b-p\hspace{-.45em}/ \delta)
\Sigma_1 \left( \frac{m_b^2}{p^2} \right)
.\eeq
The functions $\Sigma_{1,2}$ receive
 contributions of the order
${\cal O}(\alpha_s)$ from diagram 2a with $\delta=1$,
of order ${\cal O}(G_Fm_t^2)$ from diagram 2b with
 $\delta=(1-\gamma_5)$ and of order
${\cal O}(G_F\alpha_sm_t^2)$ from diagrams
 2c--2e with $\delta=(1-\gamma_5)$.
For the  calculation of the
quark selfenergies we again employ
the  hard mass procedure.
 As suggested in
\cite{GraBroGraSch90}
remaining standard scalar integrals
 may be simplified by an expansion in
$m_b^2/p^2$ around $1$. Neglecting all higher
order terms only integrals
 on the bare mass shell
need to be evaluated:
\beq\EQN{m4}
m_b^{OS}=m_b\left[1
        + \Sigma_2(1)
        - \Sigma_2(1) \left(2\Sigma_2
^{\prime}(1)+\Sigma_1(1)\right)
        + \cdots        \right]
\eeq
The derivatives
$\Sigma_2^{\prime}\equiv \partial
\Sigma_2/\partial (m_b^2/p^2)$
may be conveniently obtained through
derivations with respect to the bare
 mass, thus raising the power in the
denominator of the quark propagator.
One obtains
\beq\EQN{m5}
\ba{ll}\dsp
C_{\alpha}
& \dsp
=\frac{3}{2\epsilon}+\frac{5}{4}
+\frac{3}{2}\ln\frac{\mu^2}{m_b^2}
\\ \dsp
C_{\alpha_s}
& \dsp
=\frac{1}{\epsilon}+\frac{4}{3}
+\ln\frac{\mu^2}{m_b^2}
\\ \dsp
C_{\alpha\alpha_s}
& \dsp
=3\frac{1}{\epsilon^2}
+\frac{1}{\epsilon}
\left(
\frac{19}{4}+\frac{9}{2}\ln\frac{\mu^2}{m_t^2}
+\frac{3}{2}\ln\frac{\mu^2}{m_b^2}
\right)
\\ & \dsp
\frac{95}{8}+\frac{1}{3}\pi^2
+\frac{25}{4}\ln\frac{\mu^2}{m_t^2}
+\frac{13}{4}\ln\frac{\mu^2}{m_b^2}
\\ & \dsp
+\frac{15}{4}\ln^2\frac{\mu^2}{m_t^2}
+\frac{3}{4}\ln^2\frac{\mu^2}{m_b^2}
+\frac{3}{2}\ln\frac{\mu^2}{m_t^2}
\ln\frac{\mu^2}{m_b^2}
.\ea\eeq
A replacement of the bare bottom and top
masses
through the renormalized ones
according to eq.(\ref{m1})
in the results for the Born graph,
the electroweak
and the QCD
 corrected diagrams
leads to induced contributions
to the order ${\cal O}(G_F\alpha_s m_t^2)$.
Combined with  eq.(\ref{t1})
one obtains the  finite  result for the
decay rate.

\begin{figure}[t]
 \begin{center}
  \begin{tabular}{cc}
  \end{tabular}
  \begin{tabular}{ccc}
  \end{tabular}
  \caption{\label{figself} Diagrams for the renormalization of the
           bottom quark.}
 \end{center}
\end{figure}

\section{Discussion}
In the previous section the non-universal corrections to
the partial width $\Gamma(H\rightarrow
 b\bar{b})$ have been discussed.
Combining all contributions one obtains the following
result for these process dependent corrections:
\beq\EQN{d1}
\Delta\Gamma^{non-univ.}_
{H\to b\bar{b}}=
-6x_t
\Gamma_0 m_b^2
   \left[1
     +\frac{\alpha_s}{\pi}
\left(\frac{8}{3}-2\ln\frac{M_H^2}{m_b^2}\right)
\right]
\eeq
This result has been confirmed recently in \cite{KniSpi94}.

In addition the partial rate is also
influenced by universal corrections.
Process independent terms of order
${\cal O}(G_F\alpha_sm_t^2)$ originate
from
 $\Delta r$  due to the use of the
Fermi constant
$G_F$ instead of the electroweak coupling
constant $\alpha$ and
from the
renormalization of the Higgs vacuum. They are
given by \cite{Kni92}
\beq\EQN{d2}
\Delta\Gamma^{univ.}_
{H\to b\bar{b}}=-\Gamma_0m_b^2
 \left[
            \frac{\Pi^{WW}(0)}{M_W^2}
                             +\mbox{Re}\Pi^{HH\prime}(M_H^2)
                             +\cdots
 \right]
\left( 1+\delta_{QCD}\frac{\alpha_s}{\pi} \right)
\eeq
where
the dots indicate terms of subleading order and the QCD
correction factor\footnote{We thank the authors of
\cite{KniSpi94} for pointing out that in the earlier
 version of this paper this factor was missing in
eq.(\ref{d2}). As a consequence the interference term
 of the 1-loop QCD corrections and the 1-loop universal
electroweak corrections was erraneously not taken into
account.}
 is given by
$\delta_{QCD}=3-2\ln(M_H^2/m_b^2)$.

For the calculation of the
 the (unrenormalized)
 selfenergies $\Pi^{WW}$ and $\Pi^{HH}$
again the hard mass procedure
can be  applied.
Being interested in their real parts the
leading hard subgraphs are constituted
by the two loop diagrams themselves with
the $W$ selfenergy to be evaluated for
zero external momentum. Since the hard mass
procedure for the  Higgs vacuum
polarization is in fact an expansion of the
graph with respect to the external momentum,
the derivative Re$\Pi^{HH\prime}$ is readily
obtained from the next to leading contribution to
the power expansion.
After top mass renormalization the combination
eq.(\ref{d2}) of the $W$-- and the $H$-- selfenergy
is finite in their sum at order
${\cal O}(G_F\alpha_sm_t^2)$.
The result reads
\beq
\Delta\Gamma^{univ.}_{H\to b\bar{b}}=\Gamma_0m_b^2
   x_t
    \left[7
   -2\frac{\alpha_s}{\pi}\left(3
  +\frac{1}{3}\pi^2\right)
         \right]
\left( 1+\delta_{QCD}\frac{\alpha_s}{\pi} \right)
.\eeq
The electroweak term in the first bracket
 is in agreement with
 \cite{Kni92,DabHol92} and its  QCD contribution
reproduces the result obtained by
\cite{KniSir93} (see also \cite{GroKniWol94,Kni94}).

In the sum we arrive at the final result
for the partial decay rate
\beq
\Gamma_{H\to b\bar{b}}=
\Gamma_0m_b^2
\left\{1+x_t
      \left[1 +
     \frac{\alpha_s}{\pi}\left(
-1-\frac{2}{3}\pi^2
-2\log\frac{M_H^2}{m_b^2}
       \right)\right]
\right\}
.\eeq
The QCD correction to the electroweak
result is of comparable size as the
electroweak correction itself, but of
opposite sign.
With $m_b=4.7$ GeV, $m_t=174$ GeV
 and $M_H=60$ GeV ($120$ GeV)
the electroweak contribution
of $0.32\%$ normalized to the Born width
is combined with
a QCD corrected term of
$-0.21\%$ ($-0.24\%$).

To conclude, in this work
 we have calculated the QCD corrections
to the electroweak one-loop result for the
partial Higgs decay rate $\Gamma(H\rightarrow b\bar{b})$
in the heavy top mass limit.
For intermediate Higgs masses
electroweak corrections are
significantly reduced
by the ${\cal O}(G_F\alpha_sm_t^2)$ corrections.

\vspace{5ex}
{\bf Acknowledgments}

\noindent
 We would like to
thank K.G.Chetyrkin and J.H.K\"uhn
for helpful discussions.


\end{document}